\documentclass[preprint]{revtex4-1}
\usepackage{amsmath}
\usepackage{amssymb}
\usepackage{graphicx}
\usepackage{color}
\usepackage{esint}
\usepackage[unicode=true,pdfusetitle,
 bookmarks=true,bookmarksnumbered=false,bookmarksopen=false,
 breaklinks=false,pdfborder={0 0 1},backref=section,colorlinks=false]
 {hyperref}
\usepackage{breakurl}

\makeatletter
\@ifundefined{textcolor}{}
{%
 \definecolor{BLACK}{gray}{0}
 \definecolor{WHITE}{gray}{1}
 \definecolor{RED}{rgb}{1,0,0}
 \definecolor{GREEN}{rgb}{0,1,0}
 \definecolor{BLUE}{rgb}{0,0,1}
 \definecolor{CYAN}{cmyk}{1,0,0,0}
 \definecolor{MAGENTA}{cmyk}{0,1,0,0}
 \definecolor{YELLOW}{cmyk}{0,0,1,0}
}

\begin{document}

\title{Quasiparticle Interference, quasiparticle interactions and the origin of the charge density-wave
in 2H-NbSe$_{2}$}

\author{C. J. Arguello$^{1}$, E. P. Rosenthal$^{1}$, E. F. Andrade$^{1}$,
W. Jin$^{2}$, P. C. Yeh$^{2}$, N. Zaki$^{3}$, S. Jia$^{4}$, R.
J. Cava$^{4}$, R. M. Fernandes$^{5}$, A. J. Millis$^{1}$, T. Valla$^{6}$,
R. M. Osgood Jr.$^{3,4}$, A. N. Pasupathy$^{1}$}

\address{1. Department of Physics, Columbia University, NY, NY, 10027\\
 2. Department of Applied Physics and Applied Math, Columbia University,
NY, NY, 10027 \\
 3. Department of Electrical Engineering, Columbia University, NY,
NY, 10027\\
 4. Department of Chemistry, Princeton University, Princeton NJ 08540\\
 5. School of Physics and Astronomy, University of Minnesota, Minneapolis, MN 55455 6. Brookhaven National
Laboratory, Upton NY}

\date{\today}
\begin{abstract}
We show that a small number of intentionally introduced defects can be used as a spectroscopic tool to amplify quasiparticle interference in 2H-NbSe$_{2}$, that we measure by scanning tunneling spectroscopic imaging. We show from the momentum and energy dependence of the quasiparticle interference that Fermi surface nesting is inconsequential to charge density wave formation in 2H-NbSe$_{2}$. We demonstrate that by combining quasiparticle interference data with additional knowledge of the quasiparticle band structure from angle resolved photoemission measurements, one can extract the wavevector and energy dependence of the important electronic scattering processes thereby obtaining direct information both about the fermiology and the interactions. In 2H-NbSe$_{2}$, we use this combination to show that the important near-Fermi-surface electronic physics is dominated by the coupling of the quasiparticles to soft mode phonons at a wave vector different from the CDW ordering wave vector.
\end{abstract}

\pacs{PACS numbers: ------}

\maketitle
In many complex materials including the two dimensional cuprates, the pnictides, and the dichalcogenides  the electronic ground state may spontaneously break the translational symmetry of the lattice. Such density wave ordering can arise from Fermi surface nesting, from strong electron-electron interactions, or from interactions between the electrons and other degrees of freedom in the material, such as phonons. The driving force behind the formation of the spatially ordered states and the relationship of these states to other electronic phases such as superconductivity remains hotly debated. 

Scanning tunneling spectroscopy (STS) has emerged as a powerful technique for probing the electronic properties of such ordered states at the nanoscale \cite{Kohsaka09032007, Vershinin26032004, HudsonEW_nphys1021} due to its high energy and spatial resolution. The position dependence of the current $I$-voltage $V$ characteristics measured in STS experiments maps the energy dependent local density of states $\rho({\bf r},E)$ \cite{PhysRevB.31.805,PhysRevB.56.7656}. Correlations between the $\rho({\bf r},E)$ at different points at a given energy reveal the pattern of standing waves produced when electrons scatter off of impurities \cite{Crommie_ldos}. These quasiparticle interference (QPI) features may be analyzed to reveal information about the momentum space structure of the electronic states \cite{McElroy_QPI}. The intensity of the QPI signals as a function of energy and momentum also contains information about the electronic interactions in the material \cite{KivelsonRMP,PhysRevB.67.085102}. 

In this work, we take the ideas further, showing how impurities can be used intentionally to enhance QPI signals in STS experiments and how the combination of the enhanced QPI signals with electronic spectroscopic information available in angle-resolved photoemission (ARPES) measurements can be used to gain insight into the physics underlying electronic symmetry breaking and quasiparticle interactions. By observing the electronic response to the addition of dilute, weak impurities to the charge density wave material 2H-NbSe$_{2}$ we directly measure the dominant electronic scattering channels. We show conclusively that Fermi surface nesting does not drive CDW formation  and that the dominant quasiparticle scattering arises from soft-mode phonons.

Our theoretical analysis begins from a standard relation between the current-voltage characteristic $\frac{dI}{dV}$ at position $\mathbf{r}$ and voltage difference $V=E$ and the electron Green's function G, valid if the density of states in the tip used in the STS experiment is only weakly energy dependent
\begin{equation}
\frac{dI(\mathbf{r},E)}{dV}=\mathrm{Tr}\left[\mathbf{M}^{\mathrm{tun}}\frac{\left(\mathbf{G}(\mathbf{r},\mathbf{r},E-i\delta)-\mathbf{G}(\mathbf{r},\mathbf{r},E+i\delta)\right)}{2\pi i}\right]
\label{currenddef}
\end{equation}  
Here $\mathbf{M}$ is a combination of the tunneling matrix element and wave functions (see supplementary material); {\bf M}  and $\mathbf{G}$ are matrices in the space of band indices. 

To calculate the changes in $dI/dV$ induced by impurities we observe that in the presence of a single impurity placed at position $\mathbf{R}_a$  the electron Green's function is changed from the pure system form $G$ to $\tilde{G}$ given by
\begin{eqnarray}
 &  & \tilde{\bf G}({\bf r},{\bf r}^{\prime},E)={\bf G}({\bf r}-{\bf r}^{\prime},E)+\label{Gimp}\\
 &  & \int d{\bf r}_{1}d{\bf r}_{2}{\bf G}({\bf r}-{\bf r}_{1},E){\bf T}({\bf r}_{1}-{\bf R}_a,{\bf r}_{2}-{\bf R}_a,E){\bf G}({\bf r}_{2}-{\bf r}^{\prime},E)\nonumber 
\end{eqnarray}
Here ${\bf T}({\bf r},{\bf r}^{\prime},E)$ is the T-matrix describing electron-impurity scattering as renormalized by electron-electron interactions. It is a matrix in the space of band indices, and  we suppress spin indices, which play no role in our considerations.  

Assuming (see supplementary material) that {\bf M} is structureless (couples all band indices equally), Fourier transforming and assuming that interference between different impurities is not important gives for the impurity-induced change in the tunneling current
\begin{equation}
\delta\frac{dI(\mathbf{k},E)}{dV} =\left(\frac{1}{v}\sum_a e^{i\bf{k}\cdot\mathbf{R}_a}\right)\frac{\Phi_{\bf k}(E-i\delta)-\Phi_{\bf k}(E+i\delta)}{2\pi i}
\label{deltan_green_ajmFT} 
\end{equation}
with $v$ the volume of the systems and the scattering function of  complex argument $z$ given in the band basis in which $\mathbf{G}$ is diagonal as
\begin{equation}
\Phi_{\bf k}(z)=\sum_{nm}\int{G^{n}_{{\bf p}}(z)T^{nm}_{{\bf p},{\bf p+k}}(z)G^{m}_{{\bf p+k}}(z)d{\bf p}}
\label{Phidef}
\end{equation}
At this stage no assumption has been made about  interactions. 

From Eq.~\ref{deltan_green_ajmFT} we see that structure in $\delta dI(\mathbf{k},E)/dV$ can arise from structure in the combination $G_{\mathbf{p}}G_{\mathbf{p}+\mathbf{k}}$ of electron propagators (Fermi surface nesting) or from structure in the T-matrix, the latter arising either from properties of the impurity or from interactions involving the scattered electrons.  Combining an STS measurement with an independent determination of $G$ (for example by ARPES) allows the two physical processes to be distinguished. However, a direct analysis of Eq.~\ref{deltan_green_ajmFT} requires precise measurement of the positions of all of the impurities so that the $\sum_ae^{i\bf{k}\cdot\bf{R}_a}$ factor can be divided out. This is impractical at present, so we focus on $\left|\delta dI(\mathbf{k},E)/dV\right|$ where for dilute randomly placed impurities the prefactor can be replaced by the square root of the impurity density.  Eq.~\ref{deltan_green_ajmFT} can be further simplified if one assumes that the $T$ matrix depends primarily on the momentum transfer $k$ and has negligible imaginary part (i.e. scattering phase shift near $0$ or $\pi$). Such an assumption is particularly appropriate when the scattering arises from weakly scattering uncharged point impurities.  We find
\begin{equation}
\left|\delta\frac{dI(\mathbf{k},E)}{dV} \right|=\sqrt{n_{imp}}\left|\sum_{nm}B^{nm}({\bf k},E)T^{nm}_{{\bf k}}(E)\right|\label{QPIfromT}
\end{equation}
with 
\begin{equation}
B^{nm}({\bf k},E)=\sum_{p}\frac{G^{n}_{{\bf p}}(E-i\delta)G^{m}_{{\bf p+k}}(E-i\delta)-(\delta\leftrightarrow -\delta)}{2\pi i}\label{B}
\end{equation}
An  integral of $B$ over the occupied states yields the components of the noninteracting (Lindhard) susceptibility (see supplementary material)
\begin{equation}
\chi^{nm}_0({\bf k})\propto\int_{-\infty}^{\infty}\frac{dE}{\pi}f(E)\left(B^{nm}({\bf k},E)+B^{mn}({\bf k},E)\right)
\label{chi}
\end{equation}
where $f$ is the Fermi function.  This observation permits an interesting analysis. If the impurity scattering potential $V_{\mathrm{imp}}$ is structureless and weak,  a measurement of the QPI then directly yields the Lindhard susceptiblity. Conversely, if the impurities are known to be weak, differences between the measured QPI intensity and the Lindhard susceptibility reveal the effects of interactions, which appear formally as a ``vertex correction'' of the basic impurity-quasiparticle scattering amplitude $V_{\mathrm{imp}}$ (see supplementary material).
  
\begin{figure}
\centerline{\includegraphics[width=1\columnwidth]{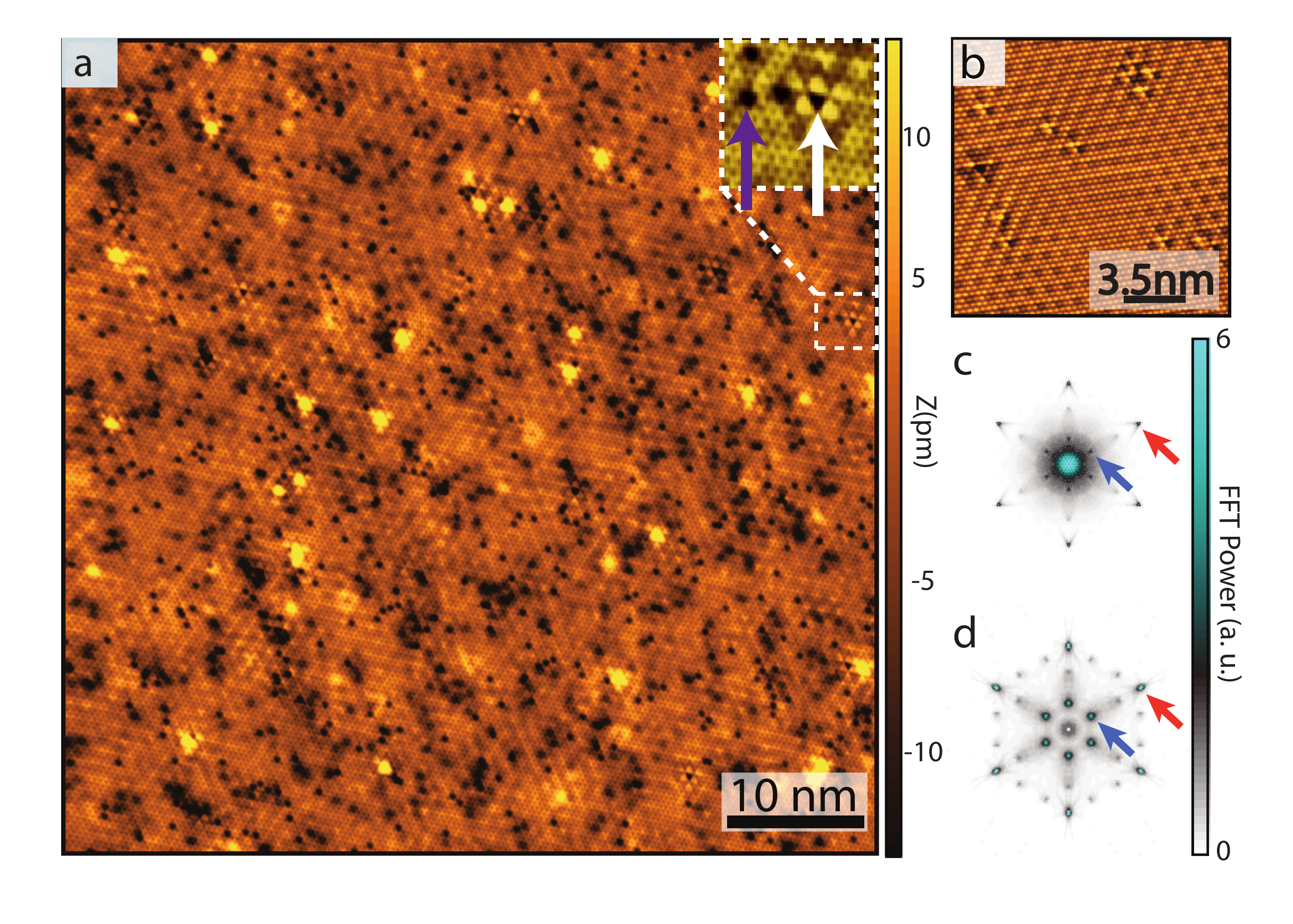}} \caption{a. Large area topographic image of NbSe$_{(2-x)}$S$_{x}$ below T$_{CDW}$, showing inhomogeneous patches of CDW. Zoomed-in region (inset) shows a sulfur dopant (purple arrow) and a vacancy (white arrow); the CDW amplitude is strongly enhanced near the vacancy. b. Topographic image of pristine NbSe$_{2}$ where the CDW is seen in all the field of view. c. Fast Fourier Transform (FFT) of the topographic image shown in (a). The inner peaks (arrow closer to origin, blue online) correspond to the CDW, the outer peaks (arrow closer to zone boundary, red online) are the atomic Bragg peaks. d. FFT of the topography of the  image  shown in (b) .}
\label{fig1} 
\end{figure}

We apply these concepts to 2H-NbSe$_{2}$, a quasi-2D transition metal dichalcogenide that displays a charge density wave (CDW) phase transition below T$_{CDW}$${\approx}$33 K\cite{PhysRevLett.34.734,1975AdPhy..24..117W, doi:10.1080/00018738800101439}. The physics of this ordered state is still under debate. While some experiments point to an important role of Fermi surface (FS) nesting \cite{PhysRevLett.82.4504,PhysRevLett.102.166402}, perhaps accompanied
by a van Hove singularity\cite{PhysRevLett.35.120,PhysRevB.63.235101}, an alternative scenario argues that the nesting of the FS is not strong enough to produce the CDW instability \cite{PhysRevB.64.235119,PhysRevB.73.205102}, and proposes that a strong electron-phonon coupling \cite{PhysRevLett.107.107403, PhysRevLett.51.138} is responsible. ARPES experiments do not detect a strong effect of the CDW order on the near-FS states.

No signatures of QPI have been detected in previous STS studies of NbSe$_2$, presumably due to the lack of sufficient scattering centers in the pristine material. To enhance the QPI signal we introduced dilute sulfur doping to pristine NbSe$_{2}$ (NbSe$_{(2-x)}$S$_{x}$). S and Se are isovalent atoms, so no charge doping arises from the substitution. Furthermore, the similarity of the calculated band structures of NbSe$_{2}$ and NbS$_{2}$ shows that the bare scattering potential induced by the substitution $\mathrm{Se}\rightarrow\mathrm{S}$ is weak. We have estimated the S-defect concentration to be approximately 1$\%$ from STM topographic images. In Fig.~\ref{fig1}(a), we show a typical topographic image taken at 27 K (T\textless T$_{CDW}$) that displays the S defects as well as a few Se vacancies. In Fig.~\ref{fig1}(b), we show a topographic image of pristine NbSe$_{2}$ in the CDW state for comparison. The CDW persists in the S-doped material as evidenced by its coverage across the entire sample, although the doped material is clearly less homogeneous than the pristine sample. This is also evident in the 2D Fast Fourier transform (2D-FFT) of the topographic images for the doped (Fig.~\ref{fig1}(c)) and pristine (Fig.~\ref{fig1}(d)) samples. Well defined CDW peaks at $\mathbf{k}{}_{CDW}$=$\mathbf{k}{}_{Bragg}/3$ are seen in the FFT for the pristine sample. These peaks broaden in the doped material, though the periodicity of the CDW does not change. Interestingly, in both materials the CDW is enhanced in the neighborhood of the Se vacancies

\begin{figure}
\centerline{\includegraphics[width=1\columnwidth]{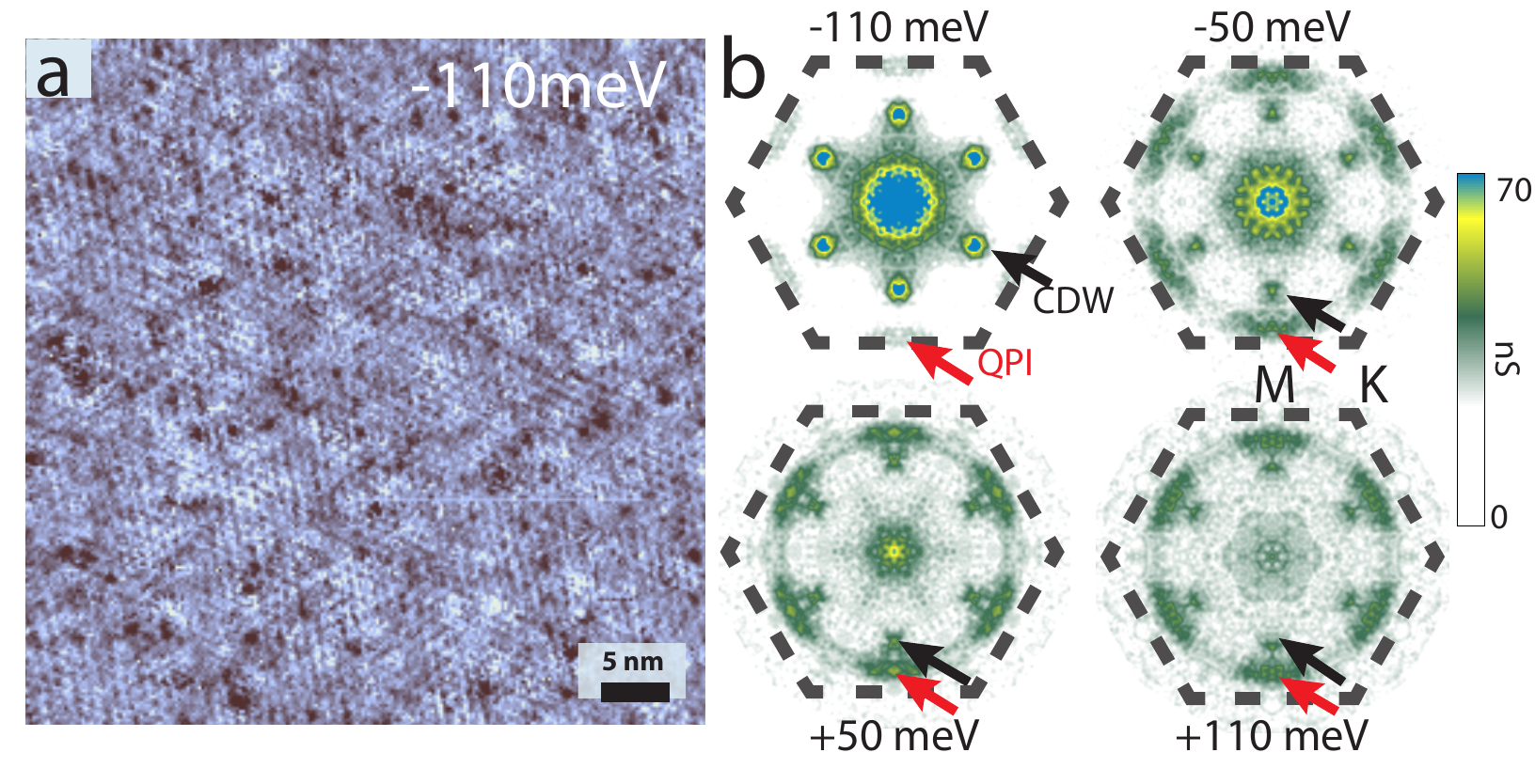}} \caption{a. Real space dI/dV map at E=-110meV. The readily visible triangular lattice arises from the charge density wave (additional real space STS images shown in the supplementary material). b. Absolute value of the FFT of dI/dV maps at different energies showing nondispersing CDW peak (heavy arrow, black online) and QPI peaks dispersing with energy (light arrow, red online). Maps have been rotationally symmetrized as described in the main text.
}
\label{fig2} 
\end{figure}

Fig.~\ref{fig2}(a) shows a typical STS map $\frac{dI}{dV}({\bf r},E)$  in real space. Fig.~\ref{fig2}(b) shows the square root of the Fourier power of the dI/dV maps, $\lvert\frac{dI}{dV}({\bf {k}},E)\lvert$ at four different energies. These Fourier transforms (FT) have been symmetrized to reflect the 6-fold symmetry of the system. Two important features are present in the Fourier transforms for all probed energies. First, there are peaks at $\mathbf{k}\simeq\mathbf{k}_{Bragg}/3$ (black arrows in Fig.~\ref{fig2}(b) at all energies measured by STS. This feature has been seen before in the pristine sample \cite{pristinenbse2} and is a consequence of the CDW order. A second feature  occurs  along the same direction as the CDW wavevector but at an energy-dependent position. Since this feature disperses in $k$ as $E$ is changed, we identify it as a QPI signal. Thus, the light doping introduced in the system successfully enhances the QPI signal while not altering the electronic structure of NbSe$_{2}$.

\begin{figure}[b]
\centerline{\includegraphics[width=1\columnwidth]{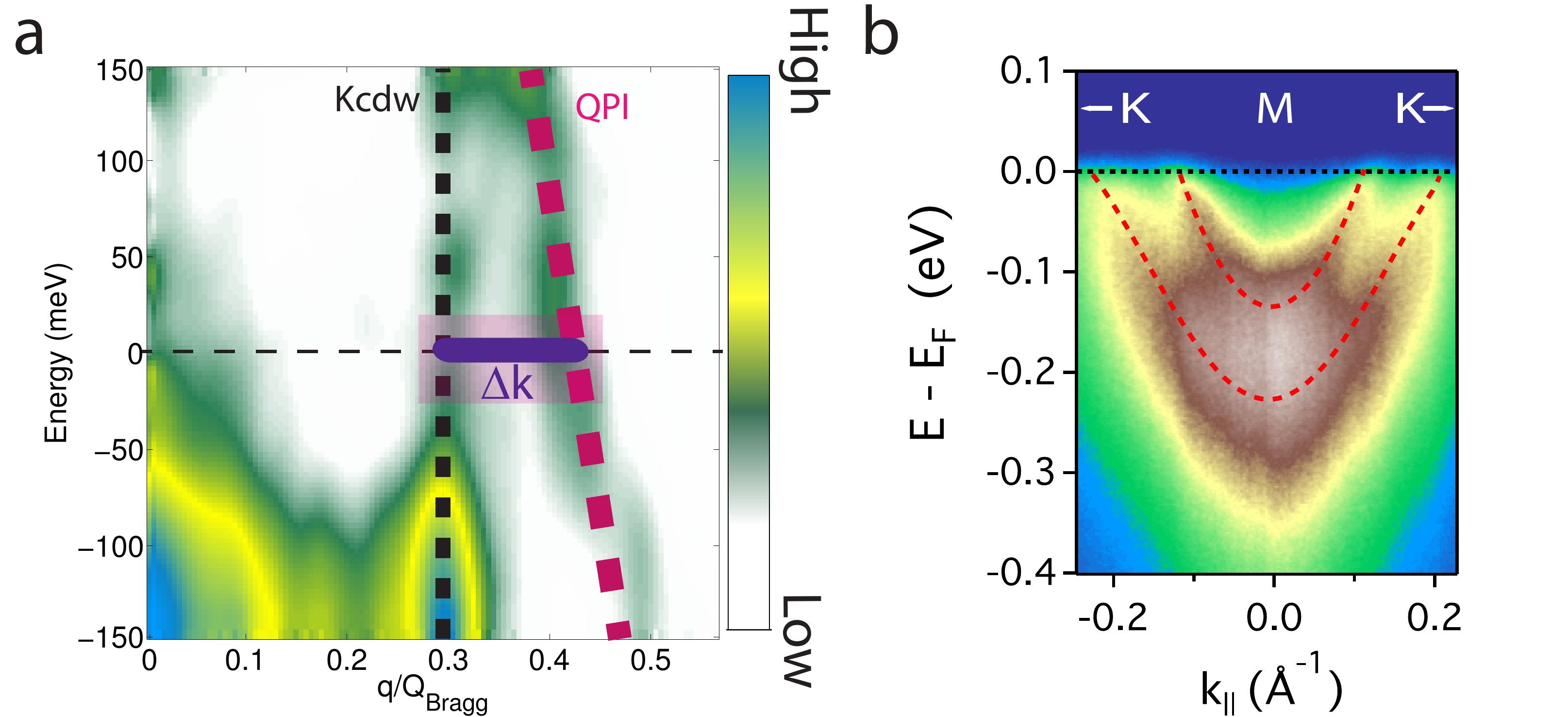}} \caption{a. Line-cut of the dI/dV maps in Fourier space along the $\Gamma-M$ direction. The dashed lines are guides for the eye highlighting the positions of the CDW ordering vector and the quasiparticle dispersion while the heavy line and shading (pink online) highlights   the separation between the QPI intensity and the CDW wavevector at the Fermi energy. (b) ARPES line-cut along the $K-M-K$ direction. The dotted line is the tight-binding fit to the data}
\label{fig3} 
\end{figure}

From Fig.~\ref{fig2}(b) we see that the QPI peaks are located at wavevectors close to the Brillouin zone edge for E=-110meV and move towards the zone center with increasing energy. We see however that for all energies presented in this paper the QPI peaks remain far from the CDW wave vector. This is illustrated more clearly in Fig.~\ref{fig3}(a) which presents a line-cut of the STS data along the $\Gamma-M$ direction for each one of the energy slices of the STS maps.  At the Fermi energy, the QPI signal is separated from the CDW signal by $\Delta\mathbf{k}\simeq\frac{1}{3}{\bf k}_{CDW}$. Extrapolation to higher energies suggests that $k_{QPI}$ would reach $k_{CDW}$ only at $E\gtrsim 300meV$ above the Fermi level.

Combining ARPES and STS measurements allows us to extract important additional information about the nature of scattering near the Fermi level in the CDW state of NbSe$_{2}$. Representative ARPES measurements are presented in Fig. ~\ref{fig3}(b). Comparison to similar data obtained on the pristine material \cite{PhysRevLett.102.166402,1367-2630-10-12-125027} revealed no significant changes in the band dispersion, further confirming that study of the lightly S doped system reveals information relevant to pristine NbSe$_{2}$. We  fit our ARPES measurements in Fig.~\ref{fig3}(b) to a two-band, five nearest-neighbor, tight-binding model similar to the one presented in Ref.~\onlinecite{PhysRevB.85.224532} (see supplementary materials for details). This fit captures the primarily two-dimensional Nb-derived bands that are believed to dominate the physics. In calculated band structures an additional Se-derived band is nearly degenerate with the Nb-derived bands near the $\Gamma$ point but disperses away as either $k_z$ or in the in-plane $k$ is increased; this band is typically not observed in ARPES experiments, most probably because of broadening associated with strong $k_z$ dispersion and for the same reason will make a much less important contribution to the QPI (see supplementary information for the details of the bands, parameters of the fit and discussion of the Se-derived states).  The fit indicates a moderate-strong (factor of $2-3$) renormalization of the observed bands relative to the calculated \cite{PhysRevB.64.235119,PhysRevB.73.205102}  bands, as previously noted \cite{PhysRevB.85.224532}. Using this band structure, we then calculate $G$ and hence $B^{mn}({\bf k},E)$ from Eq. \ref{B}.  Fig.~\ref{fig4}(a) shows the result of the calculation as well as the FT-STS measurements at the same energies (see supplementary material for comparisons between FT-STS and $B$ at additional energies). A highly structured $B$ is found, but the structures have only an indirect relation to the experimental QPI spectra. In particular, $B$  exhibits highest intensity near the K point of the Brillouin zone, where the QPI features are weak and does not exhibit significant intensity where the QPI features are strongest. 

To further characterize the differences between the quasiparticle band structure and the QPI, we  assume that the T-matrix couples all states equally ( $T^{mn}_{{\bf k}}(E)=T_{{\bf k}}(E)$ independent of band indices $mn$) and construct an experimental  estimate of $T_{{\bf k}}(E)$ from Eq. \ref{QPIfromT} by  dividing the measured $\left|\delta dI(\mathbf{k},E)/dV\right|$ by the calculated $\sum_{nm}B^{nm}({\bf k},E)$. The resulting  $|T_{{\bf k}}(E)|$ in shown in Fig.~\ref{fig4}(b) . The strong and non-dispersing peak  seen in $T_{{\bf k}}(E)$ at the CDW wave vector (indicated by the green arrows in Fig. ~\ref{fig4}(b)) is similar to the structure factors seen in X-ray diffraction experiments\cite{Chen1984645,Williams19751197}.  It is caused by the deformation of the band structure due to the periodic potential arising from the CDW ordering. Its lack of dispersion shows directly that this feature in our STS signal does not arise from quasiparticles.   

We now consider the structure highlighted as a strong peak in $T$ near the zone edge in the $\Gamma-M$ direction indicated by the purple arrows in Fig.~\ref{fig4}(b).  All available evidence suggests that the potential induced by the S-dopants is weak and structureless, so that the enhancement is an interaction effect. The strong momentum dependence of $|T|$ indicates that the intensity variation of the STS signal is not explained by the quasiparticle band structure.  However, it is significant that  at all measured energies,  the strong peak in $T$ lies within the $|\mathbf{k}|$ region delineated by the group of approximately concentric circles seen in the calculated B (denoted by the black boxes in Fig.~\ref{fig4}(a)  The main contribution to these circles arises from $2k_F$  backscattering across each of the Fermi surfaces. This suggests that the observed QPI arises from an enhancement of backscattering \cite{PhysRevB.85.104521}  by a strongly direction-dependent interaction
\cite{PhysRevLett.92.086401}. Available calculations \cite{PhysRevLett.107.107403,PhysRevB.87.245111} suggest that soft acoustic phonons with wavevector along the $\Gamma-M$ direction are strongly coupled to electrons for a wide range of $|\mathbf{k}|$. By contrast the high intensity regions in $B$ near the K point arise from approximate nesting of the Fermi surfaces centered at $\Gamma$ and $K$; that these are not seen in the measured QPI again confirms that nesting is not enhanced by interactions and is not important in this material. We therefore propose that the observed QPI signal arises from a renormalization of a structureless impurity potential  by the electron-phonon interaction.

\begin{figure}
\centerline{\includegraphics[width=1\columnwidth]{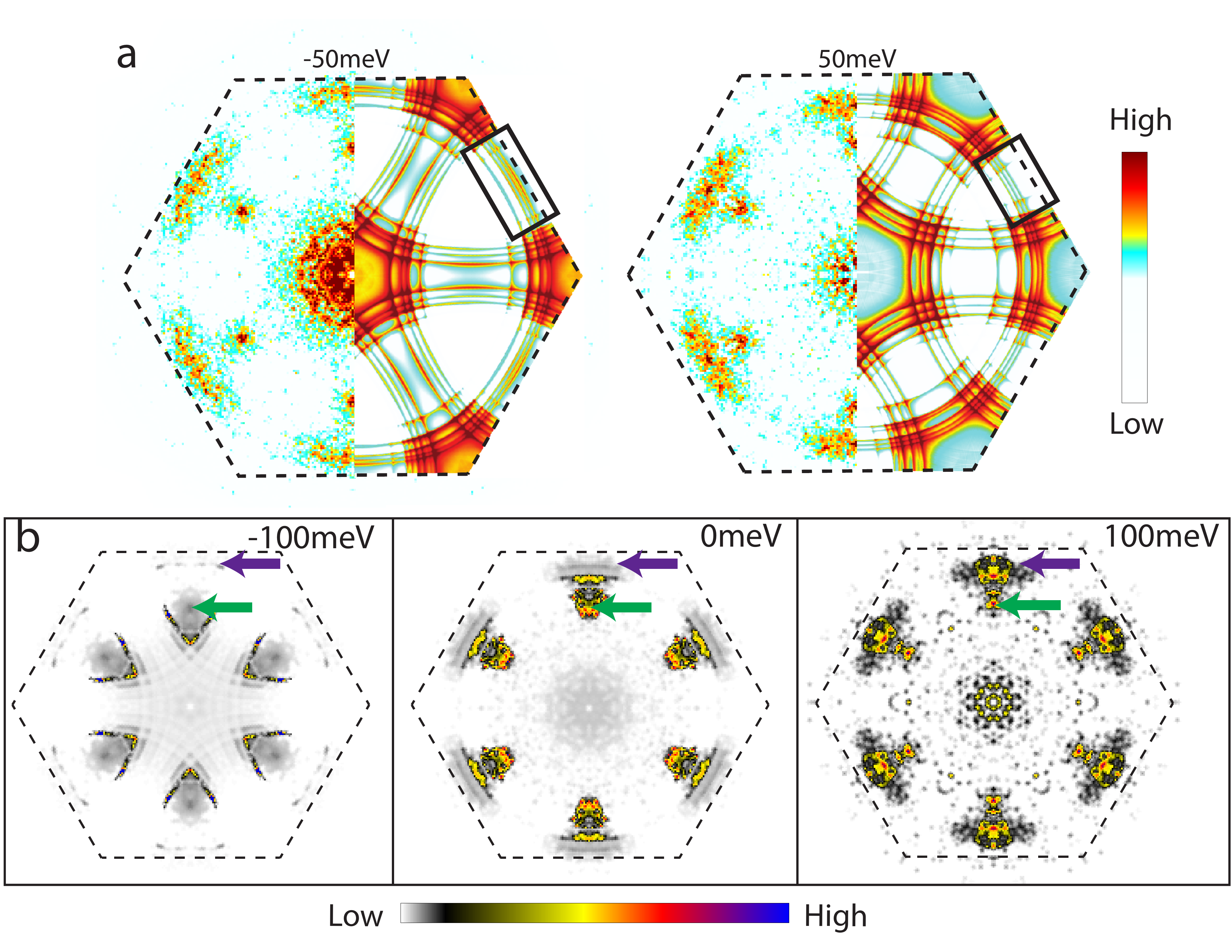}} \caption{a.  Absolute value of FFT of experimental (left half of image ) and theoretical (right half of image) dI/dV map at E=-50meV (left image) and E=50meV (right image). Theoretical images calculated as    $\sum_{mn}B^{mn}\left(\mathbf{k},E\right)$  using Eq. \ref{B} with  tight-binding model bands obtained from fits to ARPES measurements as described in the supplementary material. The dotted line is the edge of the first Brillouin zone. The black boxes indicate the areas in $\bf{k}$ space where the T-matrix is strongly peaked. b. $\left|T_{{\bf k}}(E)\right|$ calculated from the STS data using eq. \ref{QPIfromT} for -100meV (left), Fermi energy (center) and 100meV (right). The green arrow points to the position of the CDW wavevector, the purple arrow points to the 
dispersing feature in the T-matrix.}
\label{fig4} 
\end{figure}

In summary, we used dilute doping of NbSe$_{2}$ with isovalent S atoms to enhance the QPI signal and, by combining STS and ARPES measurements were able to show that the QPI signal measures more than just the fermiology of the material. We were able to confirm that the CDW does not arise from Fermi-surface nesting and we identified an important quasiparticle interaction, most likely of electron-phonon origin. Our approach reveals that the response to deliberately-induced dopants is an important spectroscopy of electronic behavior. We expect it can be extended to many other systems. 

\begin{acknowledgments}
\textbf{Acknowledgements}: \\
This work was supported by the National Science Foundation (NSF) under grant DMR-1056527 (ER, ANP) and by the Materials Interdisciplinary Research Team grant number DMR-1122594 (ANP,AJM). Salary support was also provided by the U.S. Department of Energy under Contract No. DE-FG 02-04-ER-46157 (W.J., P.C.Y., N.Z., and R.M.O.) and DE-SC0012336 (RMF). ARPES research carried out at National Synchrotron Light Source, Brookhaven National Laboratory is supported by the U.S. Department of Energy, Office of Basic Energy Sciences (DOE-BES), under the Contract No. DE-AC02-98CH10886. The crystal growth work at Princeton University was funded by DOE-BES grant DE -FG02-98ER45706. AJM acknowledges the hospitality of the Aspen Center for Physics (supported by  NSF Grant 1066293) for hospitality during the conception and writing of this manuscript. 

 \end{acknowledgments}

\bibliographystyle{apsrev}
\bibliography{references}

\newpage

{\centering{\section*{Supplementary Material for\\ ``Quasiparticle Interference, quasiparticle interactions and the origin of the charge density-wave in 2H-NbSe$_{2}$''}}}

\maketitle

\setcounter{figure}{0}

\section{Relating the charge susceptibility and QPI}
Here we present specifics of the relation between the measured QPI intensity and basic electronic properties including the bare charge susceptibility. 

\subsection{Derivation of Eq. 3 of main text} We start from the basic tunneling Hamiltonian connecting the tunneling tip to a state of the system of interest:
\begin{equation}
H_{\mathrm{tun}}= \int d^3\mathbf{r}V_{\mathrm{tun}}(\mathbf{r})\psi^\dagger_{\mathrm{tip}}\psi_{\mathrm{system}}(\mathbf{r})+H.c.
\label{Htun}
\end{equation}
Expressing the system operator at position $\mathbf{r}$ in terms of the operators $\psi_{n\mathbf{p}}$ that annihilate electrons in band state $n$ and momentum $\mathbf{p}$ in the first Brillouin zone as
\begin{equation}
\psi_{\mathrm{system}}(\mathbf{r})=\sum_{n\mathbf{p}} u_{n\mathbf{p}}(\mathbf{r})e^{-i\mathbf{p}\cdot\mathbf{r}}\psi_{n\mathbf{p}},
\label{psidef}
\end{equation}
performing the usual second order perturbative analysis of the tunneling transition rate and differentiating with respect to the voltage difference between tip and sample gives
\begin{equation}
\frac{dI}{dV}(\mathbf{r};E)=\sum_{mn;\mathbf{p}\mathbf{q}}\left|V_{\mathrm{tun}}(\mathbf{r})\right|^2u^{*}_{n\mathbf{p}}(\mathbf{r})u_{m\mathbf{q}}(\mathbf{r})e^{i\mathbf{r}\cdot\left(\mathbf{p}-\mathbf{q}\right)}\frac{G_{nm}\left(\mathbf{p},\mathbf{q};E-i\delta\right)-G_{nm}\left(\mathbf{p},\mathbf{q};E+i\delta\right)}{2\pi i}
\end{equation}
Writing a position $\mathbf{r}$ in unit cell $j$ (central position $\mathbf{R}_j$) as $\mathbf{r}=\mathbf{R}_j+\boldsymbol{\xi}$ and averaging over the in-unit cell coordinate $\boldsymbol{\xi}$ gives
\begin{equation}
\frac{dI}{dV}(j;E)=\sum_{mn;\mathbf{pq}}M^{\mathrm{tun}}_{mn;\mathbf{pq}}e^{i\mathbf{R}_j\cdot\left(\mathbf{p}-\mathbf{q}\right)}\frac{G_{nm}\left(\mathbf{p},\mathbf{q};E-i\delta\right)-G_{nm}\left(\mathbf{p},\mathbf{q};E+i\delta\right)}{2\pi i}
\label{current_1}
\end{equation}
with
\begin{equation}
M^{\mathrm{tun}}_{mn;\mathbf{pq}}=\int_{\mathrm{unit~cell}} d^3\boldsymbol{\xi}\left|V_{\mathrm{tun}}(\boldsymbol{\xi})\right|^2u^{*}_{n\mathbf{p}}(\boldsymbol{\xi})u_{m\mathbf{q}}(\boldsymbol{\xi})e^{i\boldsymbol{\xi}\cdot\left(\mathbf{p}-\mathbf{q}\right)}
\label{Mdef}
\end{equation}
Finally assuming that the combination of the  tunneling matrix element and the atomic wave functions has no interesting spatial structure ({\bf M} independent of $\mathbf{p},\mathbf{q}$)  and evaluating the momentum sums in  Eq.~\ref{current_1}   gives
\begin{equation}
\frac{dI}{dV}(\mathbf{R};E)=\sum_{mn}M^{\mathrm{tun}}_{mn}\frac{G_{nm}\left(\mathbf{R},\mathbf{R};E-i\delta\right)-G_{nm}\left(\mathbf{R},\mathbf{R};E+i\delta\right)}{2\pi i}
\label{current}
\end{equation}
which is Eq.~(1) of the main text.

\subsection{Relation between QPI and Lindhard function}
The static Lindhard or particle-hole bubble  susceptibility representing transitions between bands $n$ and $m$, $\chi^{mn}(k,\nu=0)$ may be writen
\begin{equation}
\chi(\mathbf{k})=-T\sum_{\omega_n,\mathbf{p}}\left(G^n(\mathbf{p},\omega_n)G^m(\mathbf{p}+\mathbf{k},\omega_n)+G^m(\mathbf{p},\omega_n)G^n(\mathbf{p}+\mathbf{k},\omega_n)\right)
\label{chi}
\end{equation}
Evaluating the sum in the usual way by converting to a contour integral in the complex plane which is evaluated in terms of the discontinuity across the branch cut along the real axis gives Eq.~7 of the main text. 

\begin{figure}[htbp]
\centerline{\includegraphics[width=5in]{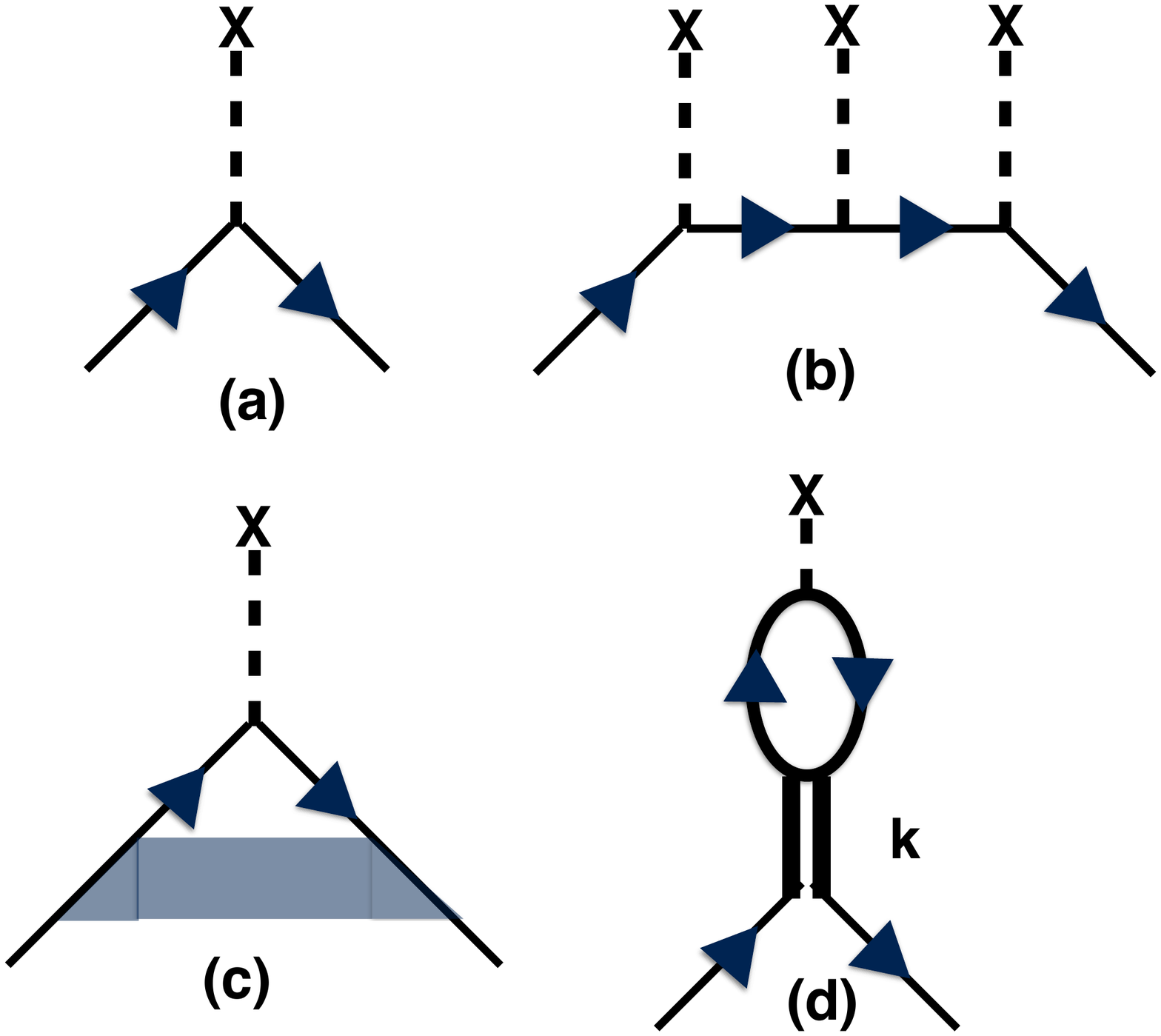}} \caption{Diagrammatic representation of electron impurity scattering. (a) single scattering event; (b) example of multiple scattering from an impurity; sum of all such diagrams yield the bare T-matrix; (c) single scattering event renormalized by general electron-electron interaction vertex (d) single scattering event renormalized by phonon. Dashed line with {\bf X}: bare electron-impurity vertex; solid line with arrow: electron propagator as determined from ARPES (i.e. renormalized by self energy); shaded box: general vertex correction; heavy double line: phonon propagator with phonon momentum $\mathbf{k}$ indicated. }
\label{S1} 
\end{figure}

\subsection{Interactions and the T-matrix}

The basic electron-impurity vertex is shown diagrammatically in  Fig.~\ref{S1}a. Multiple scattering off of the impurity is shown diagrammatically in Fig.~\ref{S1}(b). A general vertex corrections (interaction of the incoming and outgoing electron) is shown in panel Fig.~\ref{S1}(c). The particular case of an electron-phonon renormalization is shown in Fig.~\ref{S1}(d).

\section{Tight Binding Fit  to ARPES data}

Energy distribution curves (EDCs) and momentum distribution curves (MDCs) for the ARPES spectra are shown in Fig. 2(a) and  Fig. 2 (b), respectively. Quasiparticle dispersions are obtained from fits to  peak positions, which in turn are  determined by fitting the measured EDCs and MDCs to a sum of gaussians, a linear background, and a Fermi function. For example, in Fig. 2 (c), the EDC data (black dots) are fitted with two Gaussians (blue dashed curves), a linear background, and a Fermi function; in Fig. 2 (d), the MDC data (black dots) are fitted with four Gaussians (blue dashed curves) and a linear background. The energy and momentum positions of the peaks  are shown in Fig. 3 as empty circles. 

\begin{figure}
\centerline{\includegraphics[width=4 in]{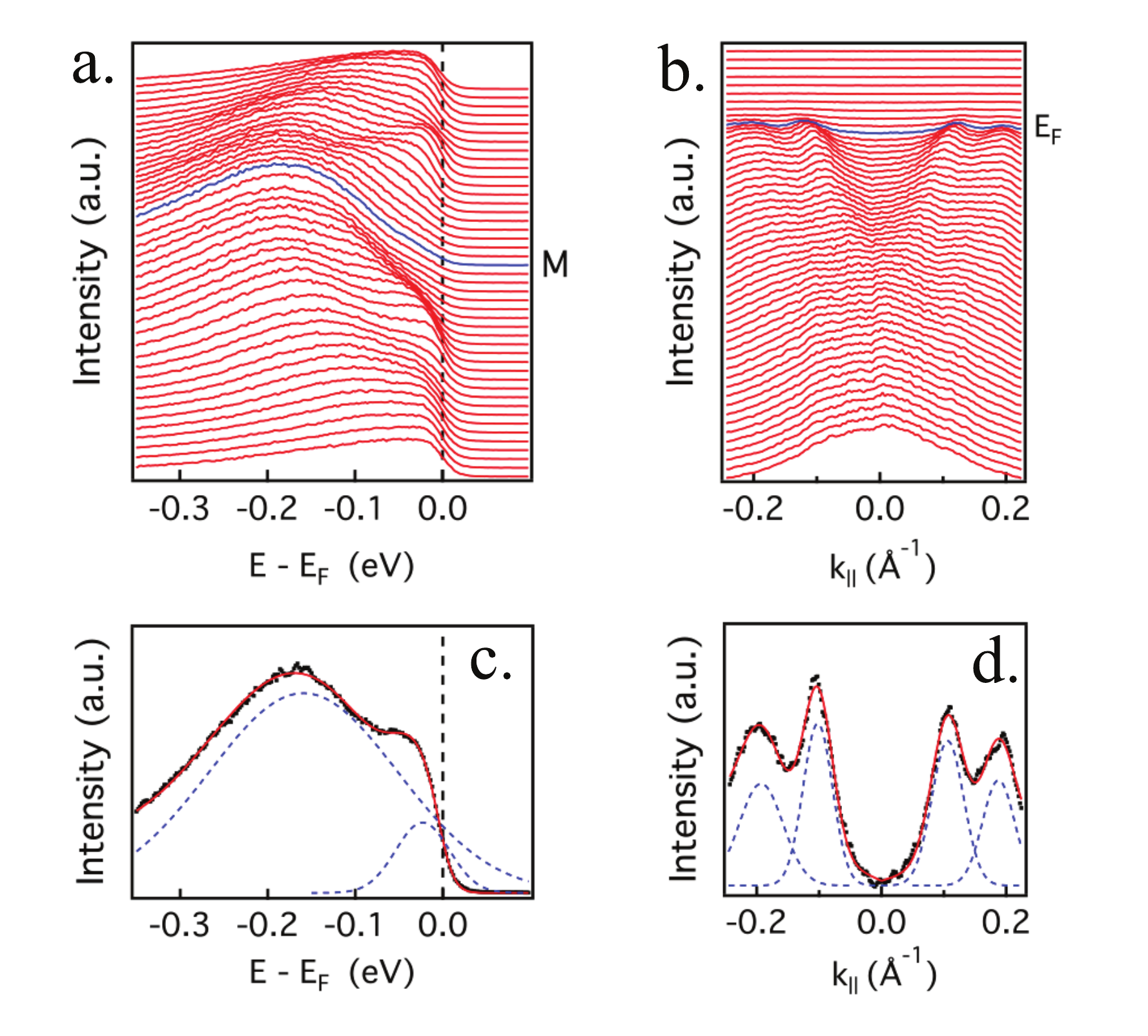}} \caption{(a) Energy distribution curve (EDCs) and (b) Momentum distribution curves (MDCs) for the ARPES spectra shown in Fig. 2 (a). (c) EDC and (d) MDC are fitted to determine the peak positions. Black dots denote the experimental data, red solid curves represent the fit curves, and blue dashed curves are Gaussians.}
\label{S2} 
\end{figure}

\begin{figure}
\centerline{\includegraphics[width=3 in]{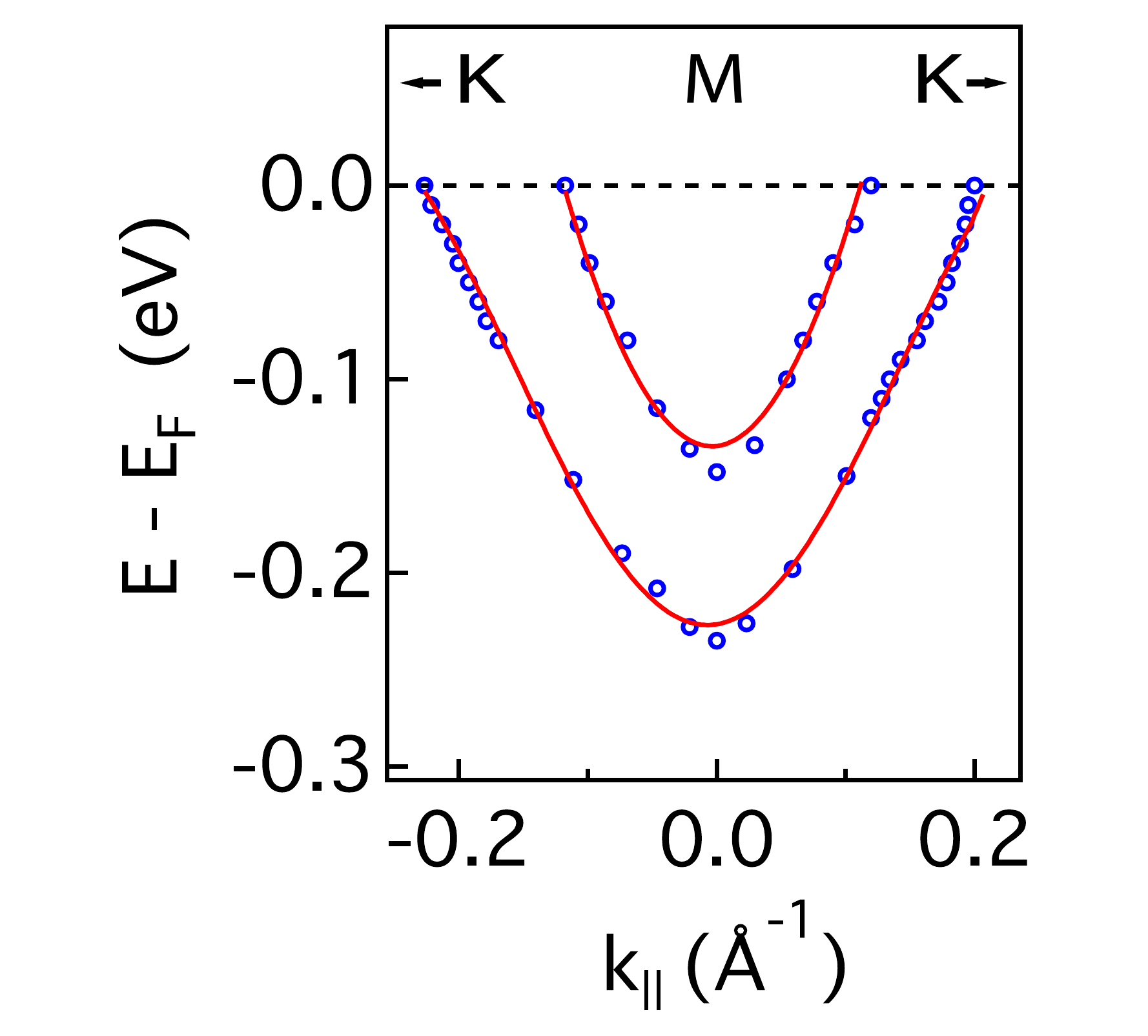}} \caption{Tight binding fit to the band dispersion. Blue empty circles denote the peak positions extracted from the EDC and MDC fitting shown in Fig. 2. Red solid curves are tight-binding fits.}
\label{S3} 
\end{figure}

We fit these two bands to a previously-proposed \cite{PhysRevB.85.224532} five-nearest-neighbor tight-binding model to extract the band dispersions (red solid curves). The bands of the tight-binding model are given by the following expression:
\begin{align}
E_i(k_x,k_y) = \nonumber  t_{0,i} &+t_{1,i} (2 \cos(\eta_{x})\cos(\eta_{y})+\cos(2\eta_{x}))\\ \nonumber &+t_{2,i} (2\cos(3\eta_{x})\cos(\eta_{y})+\cos(2\eta_{y})) \\ \nonumber &+t_{3,i}(2\cos(2\eta_{x})\cos(2\eta_{y})+\cos(4\eta_{x}))\\
\nonumber &+t_{4,i}(\cos(\eta_{x})\cos(3\eta_{y})+\cos(5\eta_{x})\cos(\eta_{y})\\ &+\cos(4\eta_{x})\cos(2\eta_{y}))  
\label{tbarpes}
\end{align}

with

\begin{equation}
\eta_x=\frac{1}{2}k_xa \qquad \eta_y=\frac{\sqrt{3}}{2}k_y a
\end{equation}

These expressions model the quasi two-dimensional Nb-derived bands that are observed in ARPES experiments. A  Se-derived band with strong $k_z$ dispersion is also found in DFT calculations \cite{PhysRevB.73.205102} but is typically not seen in ARPES \cite{Kiss_nature}. The strong $k_z$ disperson of this band also means that it will contribute less to the QPI. We do not consider it here. The parameters of the model are given in Table \ref{tbparam}. \\

\begin{table}[h!]
\begin{tabular}{|p{2.1truein}| l | l | l | l | l |}
\hline
Parameter (meV) & $t_0$ & $t_1$ & $t_2$ & $t_3$ & $t_4$ \\ \hline
Band 1 &14.2 & 82.8 & 255.4 & 42.9 & 20.5 \\ \hline
Band 2 & 265 & 21.0 & 407.2 & 8.8 & -1.0 \\
\hline
\end{tabular}
\vspace{0.2 cm}
\caption
{Tight binding parameters from ARPES. Note that in these conventions the Fermi energy is set to 0.}
\label{tbparam}
\end{table}

From the tight binding parameters we calculate the components of $B$ via the computationally efficient expression \cite{PhysRevB.73.205102}:
\begin{eqnarray}
B^{nm} ( {\bf k},E)&=& {\int_{-\infty}^0 {d\alpha \int_0^{\infty}{\frac{d \beta}{\alpha - \beta}F_{nm}(\alpha,\beta,{\bf k})}}} 
\label{Befficient1}\\ 
F_{nm}(\alpha,\beta,{\bf k})&=&\int{\frac{d{\bf k'}}{(2\pi)^2} [\delta(E_n({\bf k'})-\alpha) \delta (E_m({\bf k'}+{\bf k})-\beta)]}
\label{Befficient2}
\end{eqnarray}

\begin{figure}
\centerline{\includegraphics[width=3in]{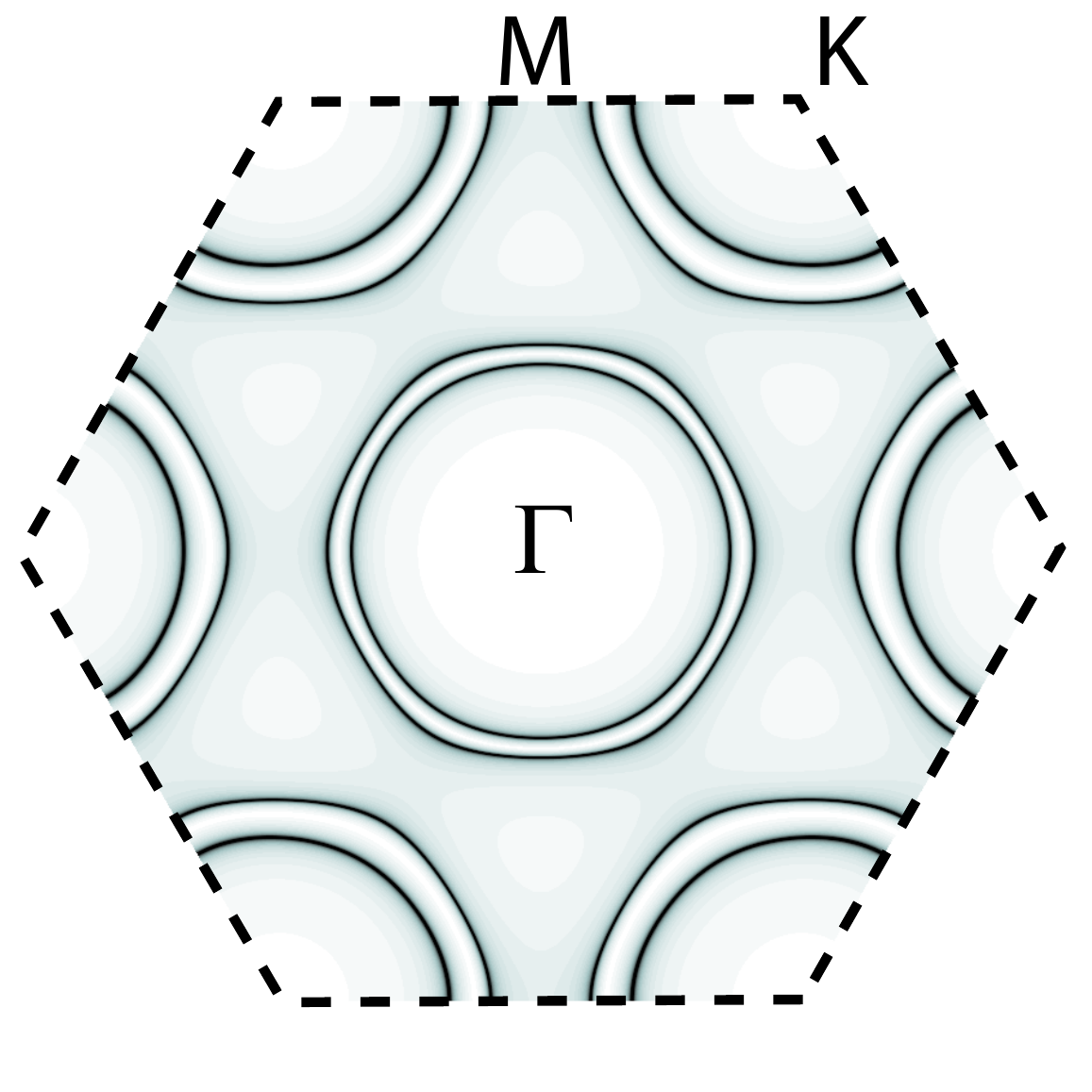}} \caption{Fermi Surface calculated from the Tight Binding 
fit to the ARPES data.}
\label{fermi_surface} 
\end{figure}

\section{Partial susceptibility calculated from STS}

Proceeding from Eq.~5  of the main text we observe that if the T matrix has negligible energy dependence and couples all bands equally then the integral of the  measured QPI signal over a range from $-E_{0} $ (chosen such that $E_{0}>>kT$) to  the Fermi level is, up to a constant, just an approximation $\chi_{0}$ to the sum of all components of the static susceptibility $\chi$:
\begin{equation}
\int^0_{- E_{0} } dE f(E) \left|\delta\frac{dI(\mathbf{k},E)}{dV} \right|\propto T(\mathbf{k}) \sum_{nm}\int^0_{-E_{0} }  dE f(E) B^{nm}({\bf k},E)= T(\mathbf{k}) \chi_{0}(\mathbf{k})\approx T(\mathbf{k}) \chi(\mathbf{k}) \label{chifromQPI}
\end{equation}

\begin{figure}
\centerline{\includegraphics[width=5 in]{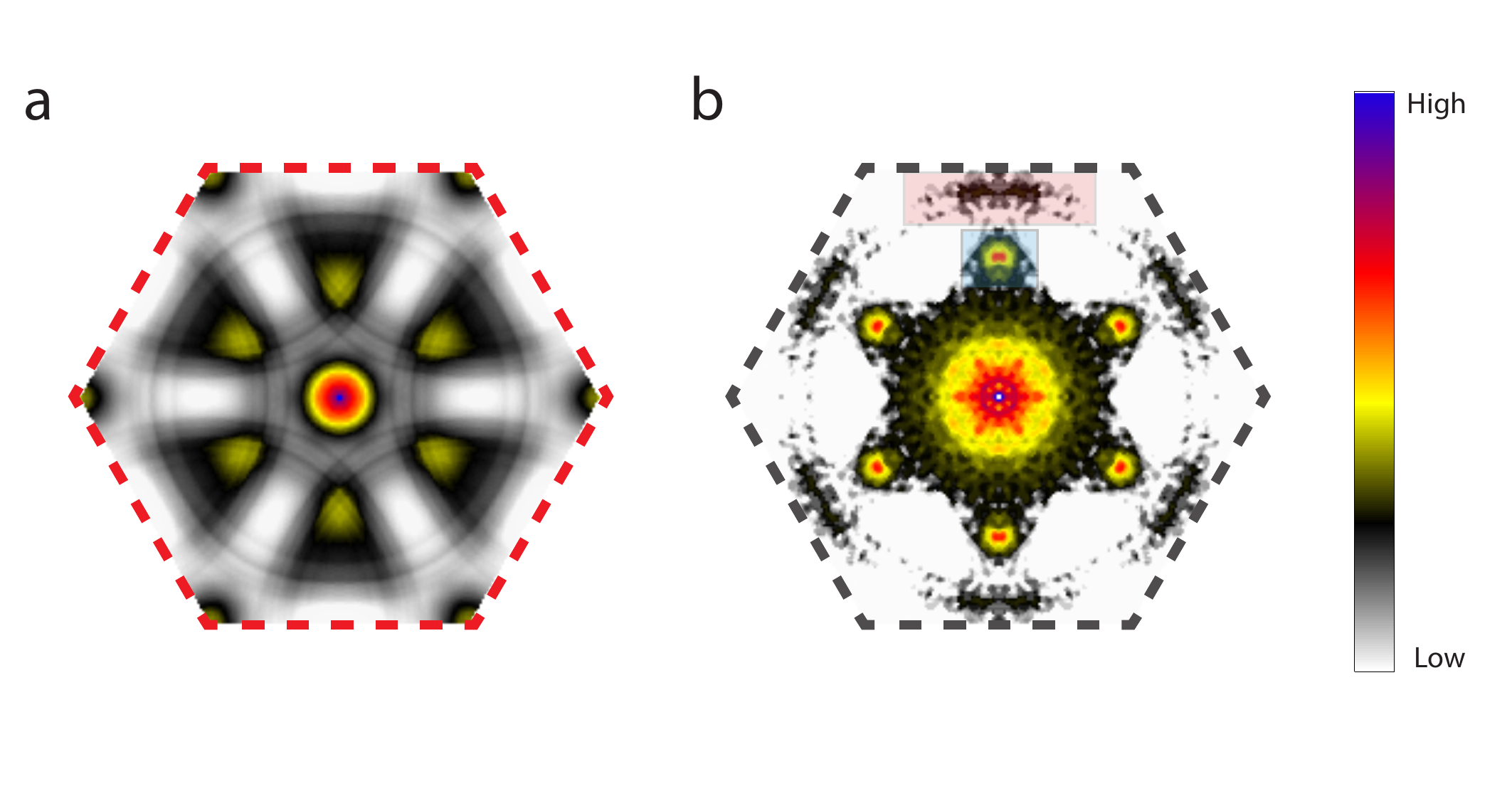}} \caption{Comparison of the electronic susceptibility calculated from ARPES (panel (a), $\sum_{mn}\chi^{mn}$ with $\chi$ from Eq ~7 of main text and $B^{mn}$ from Eqs. \ref{Befficient1},\ref{Befficient2}) and from the STS data (panel (b), Eq. ~\ref{chifromQPI}).}
\label{part_susc} 
\end{figure}

At the temperatures of the experiment (27 K), the Fermi function can be replaced with a step function, and the integral of the dI/dV signal from  $-E_{0} $ to 0 is simply the experimentally measured current $I(-E_{0}) $ .  We choose a cutoff $E_{0}=150 meV >> kT=2.5 meV$, and plot the experimentally measured  $I(-150 mV) $ in figure  \ref{part_susc}, where the portion of the signal coming from the CDW is highlighted with the blue rectangle while the dispersing QPI signal is indicated by the red rectangle. Also shown in Figure ~\ref{part_susc}  is the calculated $\chi({\bf k})$  obtained from Eq. ~7 of the text using the $B$ as computed from the ARPES bands as in Eqs.~\ref{Befficient1} and \ref{Befficient2}. The broad peaks in the $\chi$ calculated from ARPES (Fig. 5 (a)) are located at $k \simeq 0.74 k_{CDW}$ as has noted before \cite{1367-2630-10-12-125027}. The clear disagreement between the two figures points to the key role played by the momentum dependence of the T-matrix in enhancing certain scattering wave vectors in the observed QPI.

\newpage
\section{Real Space dI/dV maps}

We present here a sequence of dI/dV measurements for different energies.

\begin{figure}[htbp]
\centerline{\includegraphics[width=4 in]{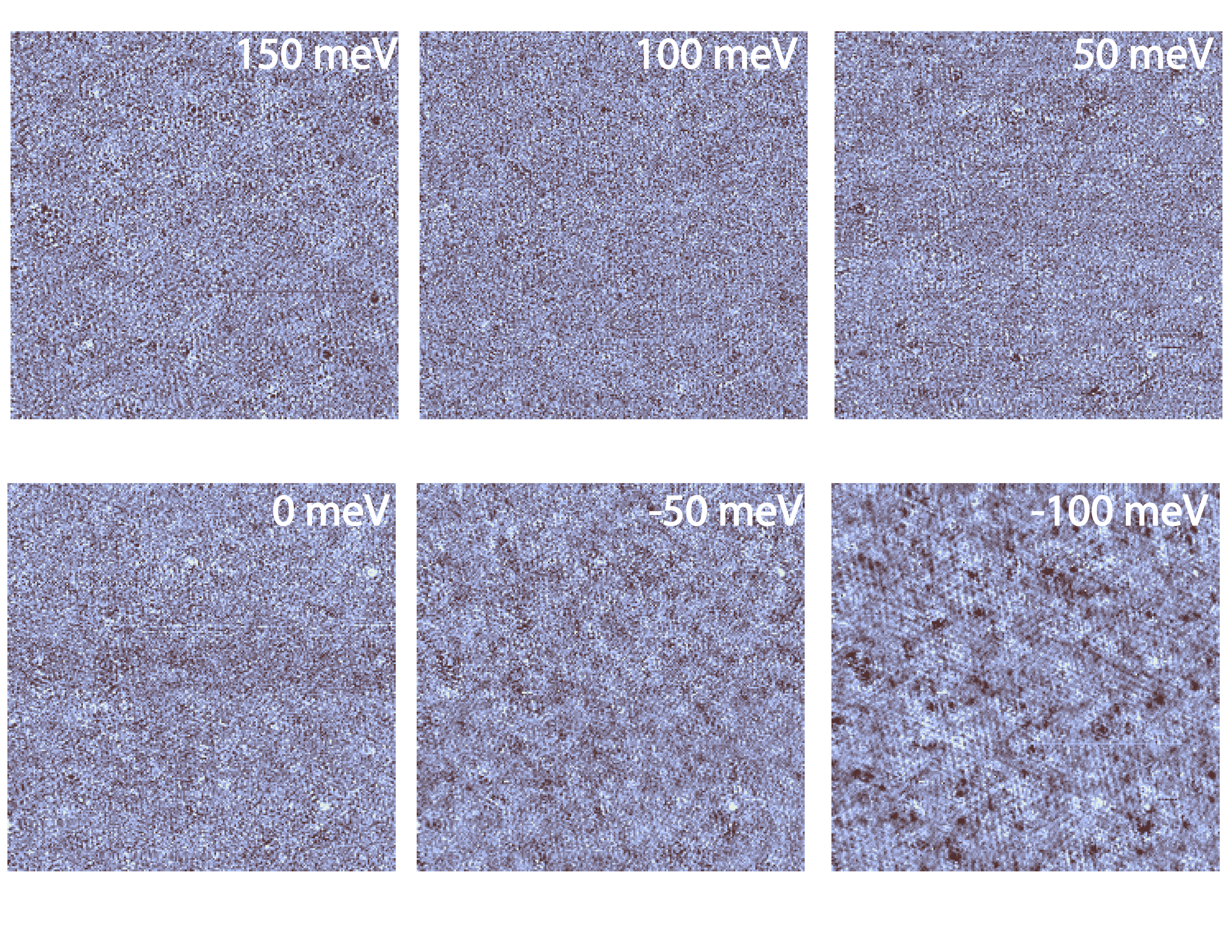}} \caption{dI/dV maps for different energies used in the main text}
\label{didv_real} 
\end{figure}

\section{Comparison between FT-STS and ARPES $B$}

\begin{figure}[htbp]
\centerline{\includegraphics[width=7 in]{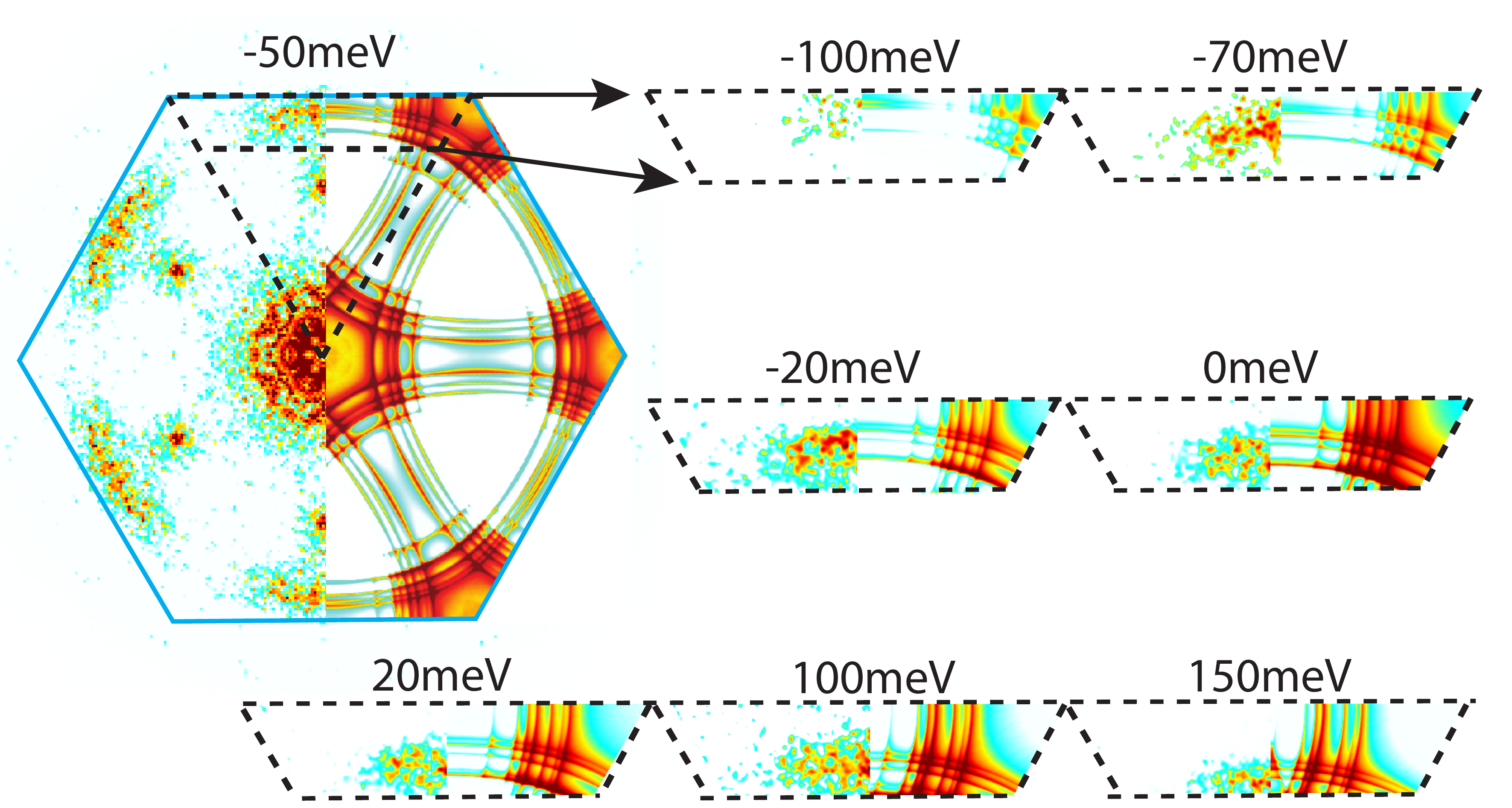}} \caption{Comparison between FT-STS and $B$ calculated from ARPES at energies indicated. Shown for each energy is a slice of k-space from the Brillouin zone boundary near the M point,  inwards to a wave vector somewhat larger than the CDW wave vector The left half of each subfigure is the FT-STS data while the right half is the calculated $B$. Within the restricted region of k-space shown, the dispersion of the FT-STS data matches very well with the $B$ calculated from ARPES at all energies. }
\label{qpiarpes} 
\end{figure}

In Figure \ref{qpiarpes} we present an expanded view ot the comparison between the Fourier transform of the measured STS data and the  $B$ calculated from ARPES at energies ranging from well below the Fermi level to well above. We zoom in a particular region of k-space that shows 
dispersing features. The left half of each subfigure is the FT-STS data while the right half is the calculated $B$. The dispersing QPI feature is located along the $\Gamma-M$ direction at wavevectors larger than $k_{CDW}$. Zooming in to the region of k-space where QPI is observed, we see from Figure \ref{qpiarpes} that the FT-STS signal is located near the edge of the Brillouin zone at energies well below $E_F$, and disperses steadily inwards at higher energies. Within this restricted region of k-space, the dispersion of the FT-STS data matches very well with the $B$ calculated from ARPES at all energies.

\end{document}